\documentclass[twocolumn,apl,amsmath,amssymb,showpacs,superscriptaddress]{revtex4-2}
\usepackage{epsf}
\usepackage[utf8]{inputenc}
\usepackage{amsmath}
\usepackage{amsfonts}
\usepackage{amssymb}
\usepackage{makeidx}
\usepackage{color}
\usepackage{gensymb}
\usepackage{graphicx}

\begin{document}
\title{Dielectric properties and impedance spectroscopy of NASICON type Na$_3$Zr$_2$Si$_2$PO$_{12}$}

\author{Ramcharan Meena}
\affiliation{Department of Physics, Indian Institute of Technology Delhi, Hauz Khas, New Delhi-110016, India}
\affiliation{Material Science Division, Inter-University Accelerator Center, Aruna Asaf Ali Road, New Delhi-110067, India}
\author{Rajendra S. Dhaka}
\email{rsdhaka@physics.iitd.ac.in}
\affiliation{Department of Physics, Indian Institute of Technology Delhi, Hauz Khas, New Delhi-110016, India}

\date{\today}      

\begin{abstract}

We report the temperature dependent dielectric properties and impedance spectroscopy investigation of Na$_3$Zr$_2$Si$_2$PO$_{12}$ in the frequency range of 20 Hz--2 MHz. The Rietveld refinement of x-ray diffraction pattern confirms the monoclinic phase with  C2/c space group. The {\it d.c.} resistivity behavior shows its strong insulating nature at low temperatures, and follows Arrhenius law of thermal conduction with an activation energy of 0.68~eV. The decrease in electric permittivity ($\epsilon_r$) with frequency is explained based on the space polarization mechanism and its increment with temperature by thermal activation of charge carriers. The dielectric loss (D=tan$\delta$) peak follows the Arrhenius law of thermal activation with an energy of 0.25~eV. We observe an enhancement in {\it a.c.} conductivity with frequency and temperature due to the decrease in the activation energy, which results in enhancing the conduction between defect states. Further, we observe an abrupt increase in the {\it a.c.} conductivity at high frequencies, which is explained using the universal Jonschers power law. The analysis of {\it a.c.} conductivity shows two types of conduction mechanisms namely correlated barrier hopping and non-overlapping small polaron tunnelling in the measured temperature range. The imaginary part of the electric modulus confirms the non-Debye type relaxation in the sample. The shifting of the relaxation peak towards higher frequency side with an increase in temperature ensures its thermally activated nature. The scaling behavior of the electric modulus shows similar type of relaxation over the measured temperature range. The combined analysis of electric modulus and impedance with frequency shows the short-range mobility of charge carriers. 
\end{abstract}

\maketitle
\section{\noindent ~Introduction}

The NASICONs known as Sodium (\textbf{Na}) \textbf{S}uper \textbf{I}onic \textbf{CON}ductors, is a class of materials having an ionic conductivity of $10^{-3}$ S/cm at room temperature (RT) and 0.2 S/cm at high temperatures ($300^\circ$C). This makes these materials as a good choice to be used as solid-electrolytes in charge storage devices such as sodium-ion batteries \cite{Xie_JPCS_16}. On the other hand, the ionic conductivity of these materials is still an order less as compared to the liquid electrolytes ($10^{-2}$ S/cm) at room temperature, and that puts a limit on using them in commercial solid-state batteries. However, various disadvantages associated with liquid electrolytes like leakage problems, flammable nature, and metallic reduction limiting their use for a longer duration, which motivates the development of solid electrolytes \cite{SunCEJ22} to have the ionic conductivity \cite{HeJMCA20} of the order of liquid electrolytes. The NASICON based materials were first discovered by Goodenough and Hong in 1976 \cite{Goodenough_MRB_76}, which show exciting structural properties depending on the chemical composition. The NASICONs having general chemical formula Na$_{1+x}$Zr$_2$Si$_x$P$_{3-x}$O$_{12}$ ($0\textless$$x$$\textless$3) show the rhombohedral structure for $x\textless$1.8 and $2.2\textless$$x$$\textless$3; whereas the monoclinic structure between $1.8\textless$$x$$\textless$2.2. The highest conductivity of 1.7$\times$10$^{-3}$ S/cm was observed for $x=$ 2, i.e., the Na$_3$Zr$_2$Si$_2$PO$_{12}$ sample \cite{Rao_SSI_21} as compared to other solid electrolytes. The structural arrangement in $x=$ 2 sample contain P/SiO$_4$ tetrahedra and ZrO$_6$ octahedra that provides a 3-dimensional network for Na ions to migrate from one place to another, giving ionic conduction \cite{Fergus_SSI_12, Xie_JPCS_16}. In both the structures (monoclinic and rhombohedral), two ZrO$_6$ octahedra are separated by three P/SiO$_4$ tetrahedra and having two different Na sites (named Na$_1$, Na$_2$). The Na$_1$ is located in between two octahedral sites along the $c$-axis and Na$_2$ is between two Na$_1$ sites along the $a$-axis, contains four Na equivalent sites per unit cell. These Na ions migrate from one site to another through oxygen triangles (formed by Na and O atoms) providing the conduction under the influence of the applied field. The conductivity depends on the size of shared triangles (also called as bottlenecks) that decides the ease of Na ion conduction; hence large conductivity is observed for large bottleneck areas \cite{Naqash_SSI_18}. 

 The Na$_3$Zr$_2$Si$_2$PO$_{\rm 12}$ has monoclinic structure at room temperature, which can be transformed to rhombohedral phase at higher temperature $\approx$ $150^\circ$C, as confirmed by various experimental techniques like {\it in-situ} high-temperature x-ray diffraction (XRD), and differential scanning calorimetry (DSC) \cite{Park_JPS_18, Jolley_JACS_15}. The variation in lattice parameters extracted using the Rietveld refinement of XRD patterns suggested that the structural phase transition at high temperatures occurs through a shear deformation of unit cell \cite{Jolley_JACS_15}. During this structural transformation, only Na ion ordering changes while other bond lengths and bond angle remain intact. The structural arrangements of both the structures show that they have four Na equivalent sites per unit cell; the rhombohedral has one Na$_1$ and three Na$_2$, out of these Na$_2$ sites are further split into one Na$_2$, and two Na$_3$ sites in monoclinic structure \cite{Park_AMI_16, Oh_AMI_19}. The high ionic conductivity and wider three-dimensional network are useful for various applications in the field of charge storage devices, like Na-ion batteries, gas sensing, microwave absorption, supercapacitors, and ion-selective electrode \cite{Guin_JPS_15, Fergus_SSI_12, Chen_JESC_18, Singh_JES_21}. There are very few studies available in literature about the dielectric and microwave absorption studies of NASICON type Na$_3$Zr$_2$Si$_2$PO$_{\rm 12}$ \cite{Jha_AIP_13,Chen_ML_18,Dubey_AEM_21}. However, these studies have not discussed the dielectric properties in detail and are limited to either high or room temperature studies. Also, the measured physical properties depend on various factors of sample preparation like the synthesis method, initial precursors, sintering conditions, 
 etc. \cite{Narayanan_SSI_19, Pal_JPCC_20,Park_AMI_16, Jolley_Ionics_15}. Therefore, it is vital to investigate the dielectric properties of Na$_3$Zr$_2$Si$_2$PO$_{\rm 12}$ at low temperatures to understand its electrical permittivity, conductivity, complex impedance, and electric modulus behavior.

In this paper, we have synthesized the Na$_3$Zr$_2$Si$_2$PO$_{12}$ bulk sample using solid-state reaction method and investigated its structural and transport properties. The phase and purity are confirmed using XRD analysis at room temperature. The {\it d.c.} resistivity is measured with temperature in two probe mode, which shows insulating nature of the sample. The electric permittivity, dielectric loss, total impedance, and phase angle are discussed with frequency at various temperatures. The {\it a.c.} conductivity and electric modulus are investigated with appropriate models to explain the relaxation dynamics and conduction behavior with temperature and frequency.

\section{\noindent ~Experimental}

 The polycrystalline bulk sample of Na$_3$Zr$_2$Si$_2$PO$_{\rm 12}$ is prepared using the conventional solid-state reaction method. All powder precursors (purity more than 99\%) named as Na$_2$CO$_3$, ZrO$_2$, NH$_4$H$_2$PO$_4$ and SiO$_2$ were preheated at 150$^\circ$C to remove any moisture. We have taken all the precursors in stoichiometric ratio with 10\% excess Na$_2$CO$_3$ to compensate for losses of Na at higher temperatures due to its volatile nature. The chemicals were mixed by grinding  thoroughly using an agate mortar pestle for uniform mixing of powders, which is initially calcinated at 1000$^\circ$C for 6 hrs in an alumina crucible. Thereafter, we perform further grinding to get the fine powder having a uniform distribution of particles and pressed into pellets of 10~mm diameter with a thickness of 2~mm using hydraulic pressure of 400 MPa. These prepared pellets are finally sintered at 1220$^\circ$C for 15 hrs with the heating and cooling rate of $5^\circ$C/min.  
 
 The powder XRD pattern was measured at room temperature using Bruker D8 advance diffractometer with Cu $K_{\alpha}$ radiation of 1.5406 \AA~wavelength. The {\it d.c.} resistivity data were collected using Keithely 6517B electrometer and Lakeshore temperature controller (Model-340) having the temperature stability of 100~mK during the measurement. We have applied 1~V excitation voltage across the sample and measured the resultant current using the LabView programme. The corrections (calibrations) were performed for open, short and cable length before starting the {\it a.c.} measurements to avoid any error due to experimental geometry. The {\it a.c.} measurements are performed using 4~T (terminal) configuration in parallel mode. For dielectric measurements using the capacitance method, we first coated both sides of the sample with silver paste and heated at $150^\circ$C for 2 hrs. The frequency dependence of the capacitance (C) and loss tangent (D=tan$\delta$) were measured as a function of frequency (100~Hz--2~MHz) and temperature with 1~V {\it a.c.} input perturbation using Agilent LCR meter (Model E4980A) and lakeshore temperature computer (model 340). We have also measured the total impedance ($Z^*$=$Z'$+j$Z''$) and phase angle ($\theta$) in the temperature range of 200--400~K within the frequency range of 20~Hz--2~MHz. The holder was placed in liquid nitrogen based dipstick bath for uniform cooling across the sample maintaining the vacuum of $10^{-2}$ mbar.

\begin{figure}[h]
\includegraphics[width=3.45in]{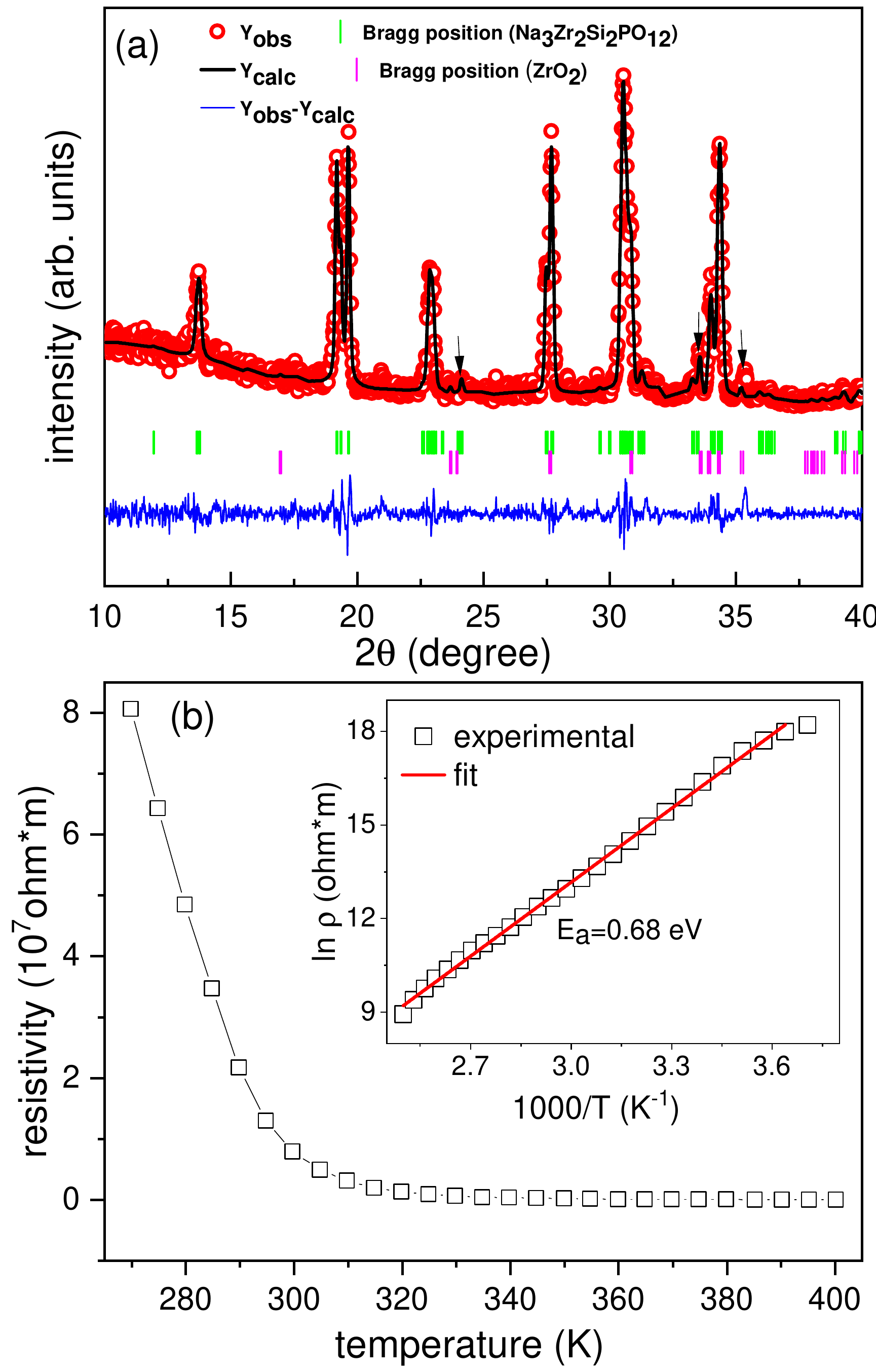}
\caption {(a) The room temperature XRD pattern (red circle) of Na$_3$Zr$_2$Si$_2$PO$_{\rm 12}$ sample.The solid black line shows the Rietveld refinement profile; green and pink short verticals show the Bragg peak positions for sample and small impurity of ZrO$_2$, and the blue line shows the residual. (b) The electrical resistivity in the temperature range of 270--400~K. The inset shows ln($\rho$) vs. (1000/T) plot, where solid line represents the linear fit of Arrhenius thermal conduction model.}
 \label{XR}
\end{figure}
	
\section{\noindent ~Results and discussion}

The XRD pattern of polycrystalline Na$_3$Zr$_2$Si$_2$PO$_{\rm 12}$ bulk sample is shown in Fig.~\ref{XR}(a) along with the Rietveld refinement using the Pseudo-Voigt function and taking a linear interpolation in the background. The refinement results confirm the monoclinic phase [space group (C 2/c)] and show a good fit to the experimentally observed XRD pattern with a reduced-$\chi^2$ value of 1.72. A small amount of impurity is visible, as denoted by arrows, due to unreacted ZrO$_2$ material  \cite{Dubey_AEM_21,Oh_AMI_19}. 
The obtained unit cell lattice parameters are $a=$ 15.655 \AA, $b=$ 9.037 \AA, $c=$ 9.178 \AA~ and $\beta$=123.965$^{\circ}$, which are in close agreement with the values reported in the literature \cite{Lalere_JPS_14}. Figure~\ref{XR}(b) shows the temperature-dependent resistivity of Na$_3$Zr$_2$Si$_2$PO$_{\rm 12}$ sample in the range of 270--400~K. The resistivity behavior indicates the strong insulating nature of the sample at low temperatures. The decrease in resistivity with temperature is due to the increased mobility of thermally induced charge carriers. We estimate the activation energy of thermally induced charge carriers using the Arrhenius model of thermal activation; according to this model, the resistivity data can be fitted using the relation
\begin{equation}
\rho_{T}=\rho_{0}~exp(\frac{E_a}{{k_B} T})
\end{equation}
here $E_a$ is the activation energy of charge carriers during thermal conduction, $k_B$ is the Boltzmann constant, T is the temperature, $\rho_0$ is the pre-exponential factor, and $\rho_T$ is the resistivity of the sample measured at finite temperature. We plot the curve between ln$\rho_T$ versus (1000/T) [inset of Fig.~\ref{XR}(b)] and used the slope to calculate the activation energy, which is found to be 0.68~eV. 

The dielectric measurements give useful information about the relaxation mechanism inside the material with its frequency and temperature dependence. It provides information about two important physical parameters of a material (a) electric permittivity (dielectric constant) and (b) dielectric loss. The dielectric constant ($\epsilon_r$) relates to the charge storage capacity of the material under the influence of applied electric field. When an alternating field is applied across the sample, the charge carriers produce the heat due to lag of polarization called dielectric loss (D=tan$\delta$). The dielectric loss inside material is the combined effects of dielectric relaxation and electrical conduction. The electric permittivity changes with temperature, frequency, orientation, pressure and molecular structure of materials. The total dielectric constant or electric permittivity ($\epsilon_r$), its real ($\epsilon_r'$) and  imaginary ($\epsilon_r''$) parts with loss tangent (D) can be calculated using the following expressions:
\begin{equation}
    \epsilon_r(\omega)=\frac{C_P}{C_0}=\frac{C_P d}{\epsilon_0 A}
\end{equation}
\begin{subequations}
  \begin{equation}
      \epsilon_r'(\omega)= \lvert \epsilon_r(\omega) \rvert  Cos\delta = \lvert \epsilon_r(\omega) \rvert Sin\theta
  \end{equation}
  \begin{equation}
      \epsilon_r''(\omega)=\lvert \epsilon_r(\omega) \rvert  Sin\delta = \lvert \epsilon_r(\omega) \rvert Cos\theta
  \end{equation}  
\end{subequations}
 \begin{equation}
      D=\tan\delta =\frac{\epsilon_r''(\omega)}{\epsilon_r'(\omega)}
  \end{equation}
here $C_P$ is the measured parallel plate capacitance, $C_0$ is the vacuum capacitance given by $C_0$=$\epsilon_0$A/d where $\epsilon_0$ is the permittivity of free space ($8.85 \times 10^{-12}$ F/m),  $A$ is the cross-sectional area, $d$ is the thickness of dielectric material, $\omega$ (=2$\pi$$f$) is the angular frequency of the applied field, and $\delta$=(90-$\theta$) where $\theta$ is the phase difference between the applied alternating voltage to the measured current.  

\begin{figure}[h]
\includegraphics[width=3.45in]{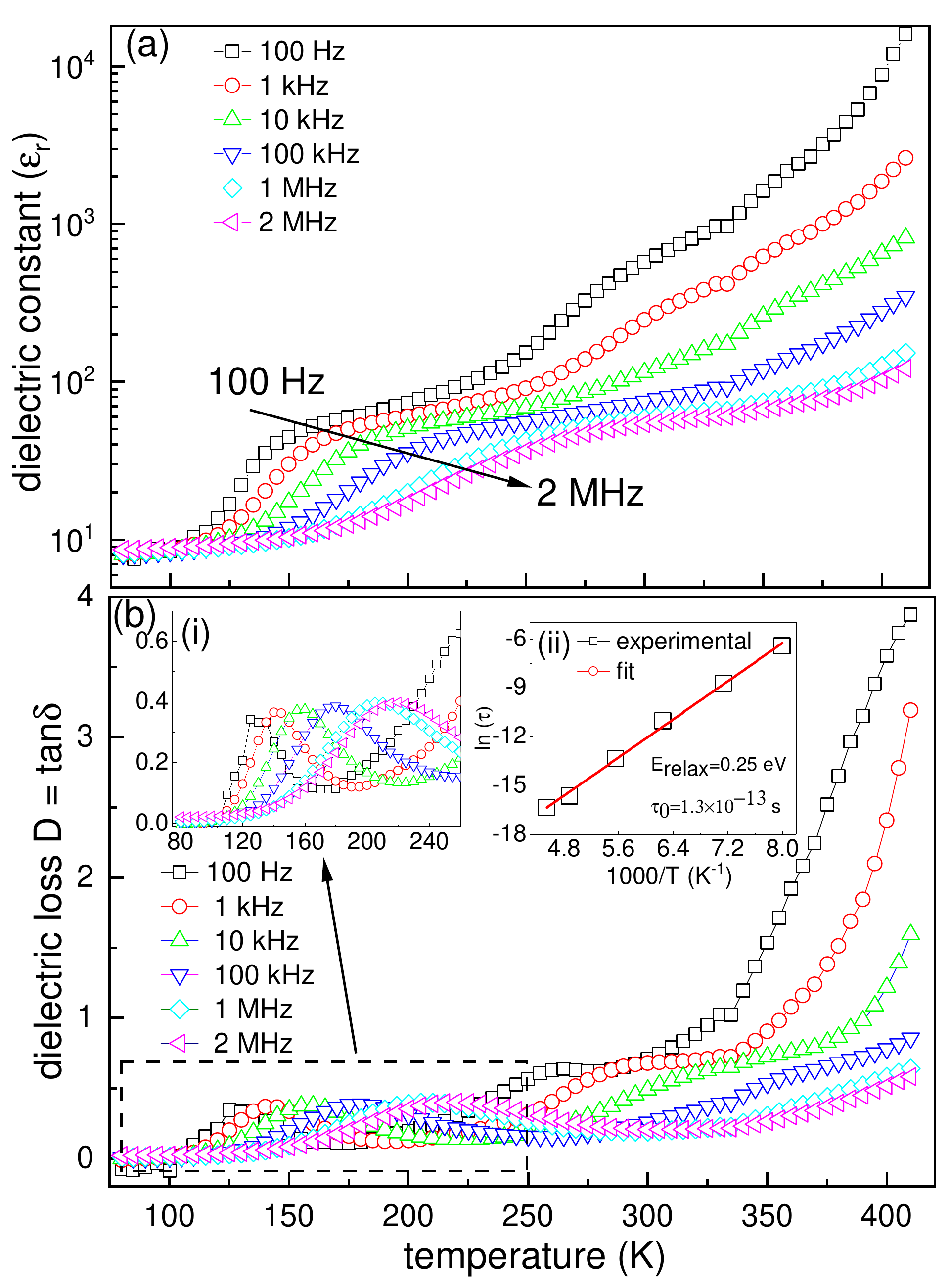}
\caption {(a) The dielectric constant and (b) loss tangent as a function of temperature for Na$_3$Zr$_2$Si$_2$PO$_{\rm 12}$ sample at different selected frequencies. The arrow in (a) shows the dielectric relaxation behavior and insets in (b) show (i) the shift of loss peak towards high temperature with an increase in frequency and (ii) the Arrhenius plot (ln($\tau$) vs. 1000/T) of relaxation time to find the activation energy of dipole relaxation.} 
\label{CDT}
\end{figure}

 The temperature dependence of dielectric constant $\epsilon_r(\omega)$ at selected frequencies is shown in Fig.~\ref{CDT}(a). The relaxation-type behavior shifts towards higher temperatures (shown by the arrow) with increasing frequency. The dielectric relaxation in materials occurs due to defects such as vacancies and space charge electrons. The thermally induced activation of charge carriers enhance the polarization and that leads to the increase in dielectric constant with temperature. The large increment in dielectric constant at lower frequencies is explained based on the space charge polarization mechanism; according to this model the charge carriers follow the applied field at lower frequency values giving an enhanced polarization results in higher values of permittivity. While, at higher frequencies the carriers do not follow the applied field completely as the  time period is much shorter, i.e., the applied field changes its direction before the carriers align in field direction; therefore, the net polarization decreases and hence dielectric constant. The total dielectric constant is the combined effect of (i) frequency of the applied field and (ii) the sample temperature. The dielectric loss variation with temperature at selected frequencies is shown in the Fig.~\ref{CDT}(b). The dielectric loss mainly depends on the temperature and mobility of charge carriers. The mobility of charge carriers increases with  temperature, resulting in an increased dielectric losses. A peak in dielectric loss data called as loss peak appears when the frequency of applied field is equal to the hopping frequency of charge carriers. The loss peak shifts towards higher temperature [as shown in the inset (i) of Fig.~\ref{CDT}(b)] with frequency due to an increase in hopping frequency of charge carriers with temperature \cite{Kolte_AIP_15, Sumi_Jap_10}. The maxima of the peak satisfied the condition {$\omega$$\tau$=1} where $\tau$ is the relaxation time of charge carriers. The shifting of loss peak towards higher temperature shows the relaxation behavior inside the material due to a thermally activated process. The activation energy of relaxation can be estimated using the Arrhenius model; according to this model, $\tau$ should follow the relation given by \cite{AngPRB00}:
\begin{equation}
     \tau = \tau_0 ~ exp(\frac{E_{relax}}{k_B T})
\label{tr}
 \end{equation}
here, $\tau_0$ is the characteristic relaxation time of the order of atom vibrational period, and $E_{relax}$ is the activation energy of relaxation. The linear plot between ln$\tau$ versus 1000/T is obtained using equation \ref{tr} [as shown in the inset (ii) of Fig.~\ref{CDT}(b)] gives the activation energy  $E_{relax}$=0.25~eV. The characteristic relaxation time $\tau_0$ found from the intercept of the curve is 1.3 $\times$ $10^{-13}$ sec.

\begin{figure} 
\includegraphics[width=3.45in]{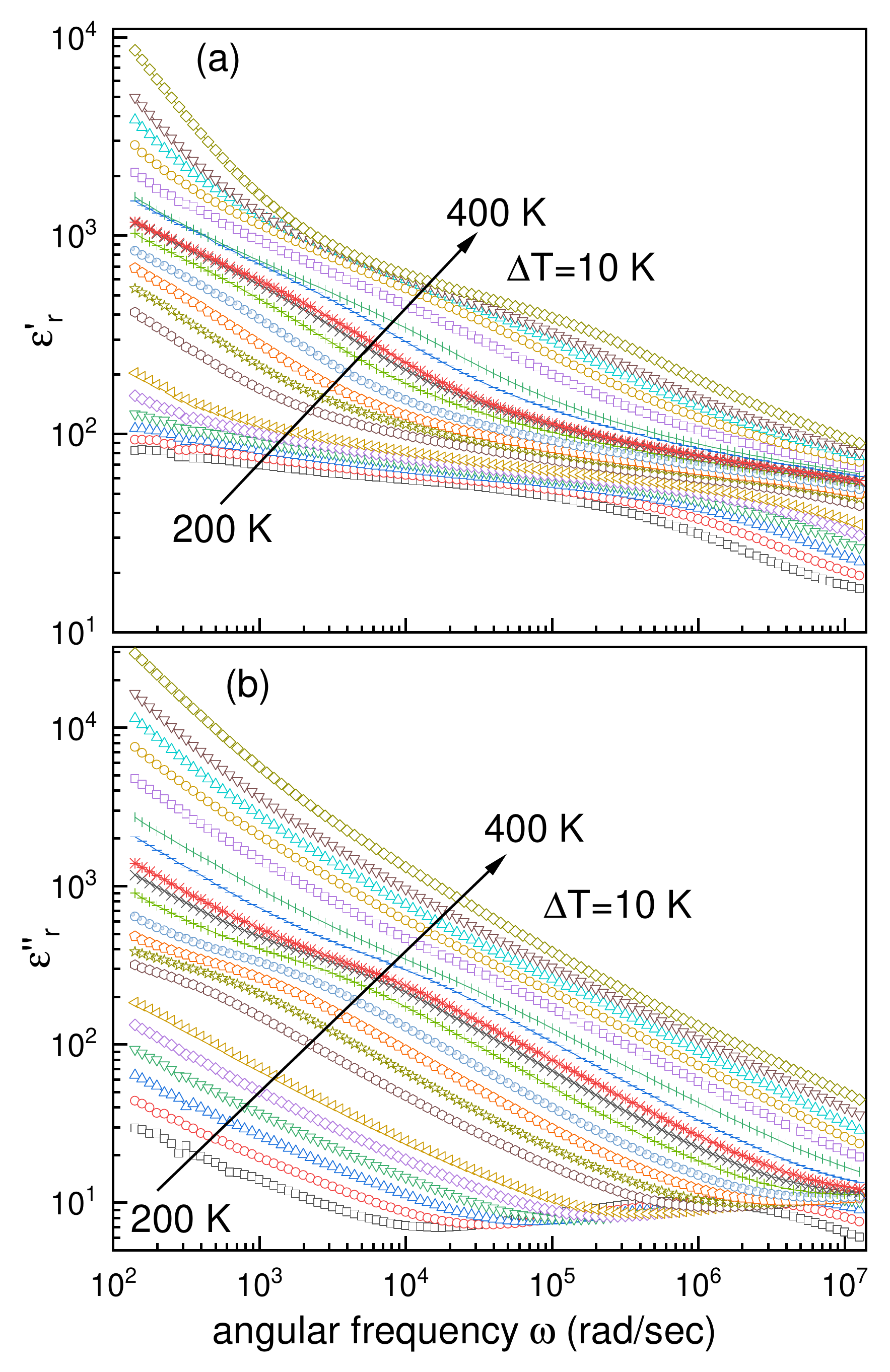}
\caption {The real (a) and imaginary (b) parts of dielectric constant as a function of frequency (20 Hz--2 MHz) at different temperature (200--400~K) for the Na$_3$Zr$_2$Si$_2$PO$_{\rm 12}$ sample.} 
\label{EET}
\end{figure}

The variation of real ($\epsilon_r '$) and imaginary ($\epsilon_r''$) parts of dielectric constant (electric permittivity) with frequency at different temperatures are shown in Figs.~\ref{EET}(a) and \ref{EET}(b), respectively. The highly dispersed region with larger permittivity values is observed at lower frequencies, indicating the hopping type conduction mechanism in the sample. The larger values of permittivity at lower frequency is explained by the Maxwell-Wagner relaxation model and space charge polarization; according to this model, a dielectric material can be considered as combination of the larger number of well-conducting grains separated by poorly conducting grain boundary regions. Under the influence of the externally applied field, the charge carriers easily migrate through the grains and accumulate at grain boundaries. The accumulation of charge carriers at the grain boundary leads to higher values of the dielectric constant \cite{Kolte_AIP_15, Sumi_Jap_10}. Considering the effects of temperature on dielectric materials, the charge carriers follow the direction of applied field at higher temperatures (as the thermal energy available with carriers helps in alignment), creating the piling of charges by successfully hopping from the lower energy barrier side to a high energy barrier. The piling of charges increase the polarization; hence an increase in permittivity \cite{Nallamuthu_JALCOM_11}. 

\begin{figure}[h] 
\includegraphics[width=3.45in]{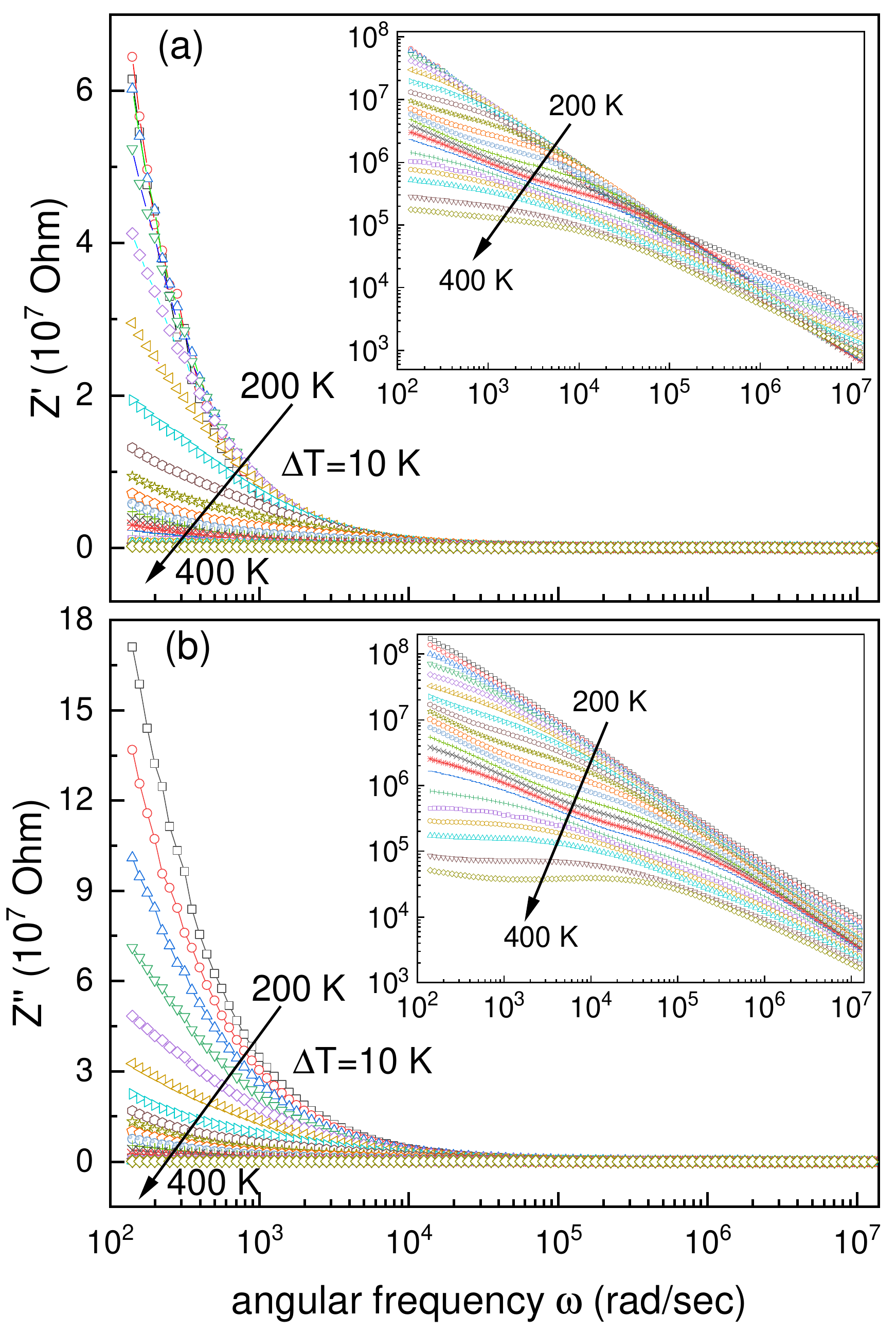}
\caption {The real (a) and imaginary (b) parts of impedance as a function of frequency (20 Hz--2 MHz) at different temperatures (200--400~K) for the Na$_3$Zr$_2$Si$_2$PO$_{\rm 12}$ sample. The insets show the same data, but in the log scale for clarity.} 
\label{ZZT}
\end{figure}

The complex behavior of total impedance with frequency can be used to analyze various electrical properties like the effect of grain and grain boundary on the total impedance, relaxation time distribution, and type of relaxation mechanism \cite{Joshi_MRB_17, Ksentini_APA_20}. The total complex impedance can be expressed using the formula 
\begin{equation}
Z^* = Z' +j Z''
 \end{equation}
where, the real $(Z')$ and imaginary $(Z'')$ parts of total impedance are given by
\begin{equation}
Z'=  \lvert Z \rvert Cos \theta  \hskip0.5cm   {\rm and}   \hskip0.5cm    Z''= \lvert Z \rvert  Sin\theta
 \end{equation}  
here, $\lvert Z \rvert$ is total measured impedance, and $\theta$ is the measured phase angle (in radian). The variation of the real and imaginary parts of impedance as a function of frequency at different temperatures are presented in Figs.~\ref{ZZT}(a) and \ref{ZZT}(b), respectively. The impedance data show a monotonic decrease for all measured temperatures indicating the negative temperature coefficient of resistance. The impedance values are high at a lower frequency and decrease at higher frequencies. The decrease in impedance is due to the release of immobile charges and increased mobility with increasing frequency. The decrease in barrier height with temperature results in lower values of impedance. The tendency of merging all the curves at higher frequencies is due to the reduction in space charge and cancellation of the dipole orientation effect. The absence of any relaxation peak shows the distribution of relaxation times corresponds to non Debye type behavior \cite{Joshi_MRB_17, Taher_MRB_16, Abdessalem_JALCOM_19}.

\begin{figure}[h]
\includegraphics[width=3.4in]{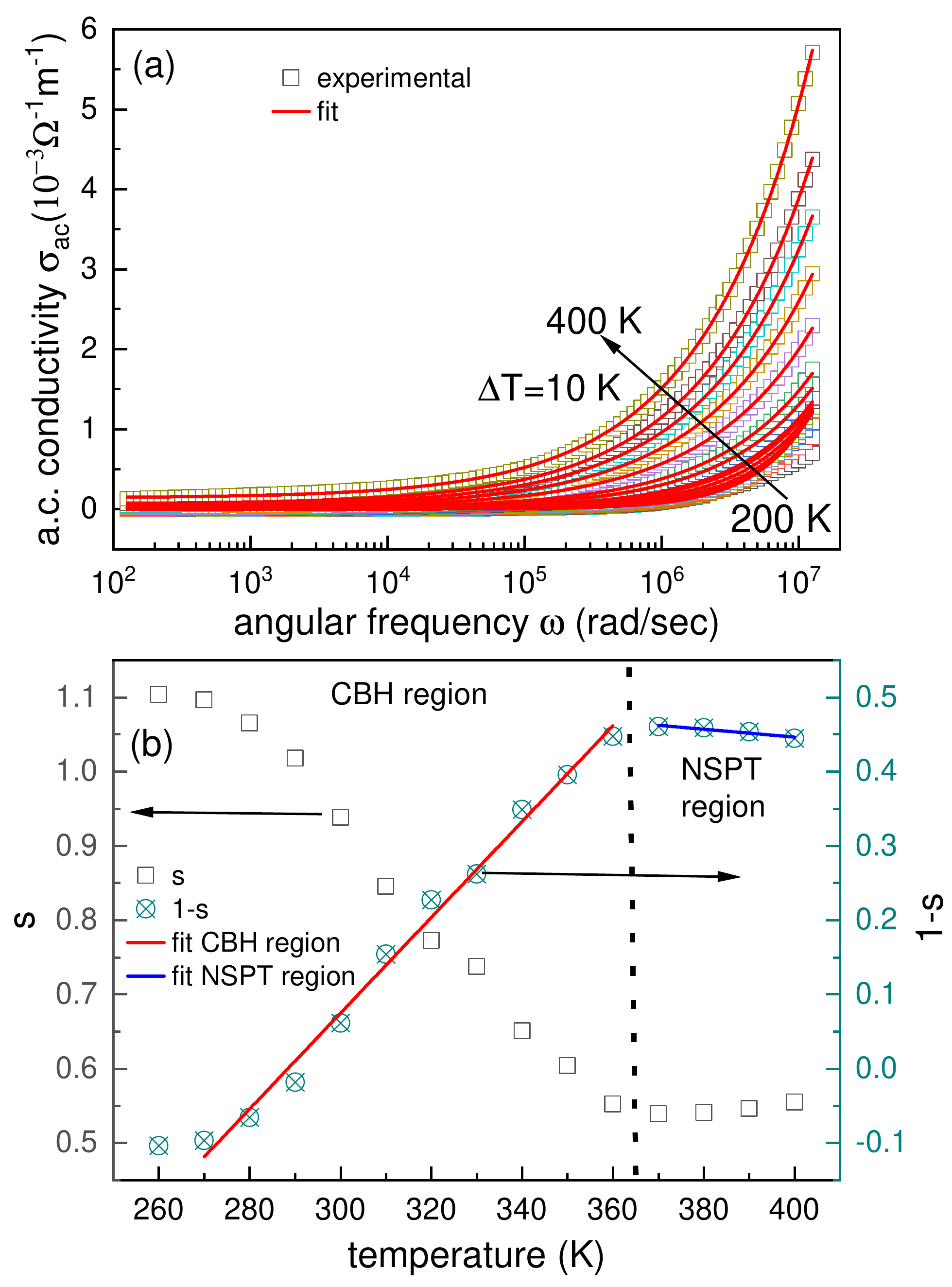}
\caption {(a) The frequency dependence of {\it a.c.} conductivity for the Na$_3$Zr$_2$Si$_2$PO$_{\rm 12}$ sample at various temperatures, and (b) the temperature dependence of $s$ parameter on left scale and (1-s) on right scale . Here, the open symbols represent the experimental data and the solid line represents the fit. The dashed line in (b) around 365~K separates the CBH and the NSPT conduction regions.} 
\label{ACS}
\end{figure}

The electrical conduction in an alternating field is contribution from carriers in phase with the applied field. The total conductivity is the contribution from free carriers and bound charge carriers. It has been reported that if the conductivity decreases with an increase in frequency, the dominating contribution is from free charge carriers; whereas, if it increases with frequency, the conductivity is dominated due to bound charge carriers \cite{Sumi_Jap_10}. The alternating field across the sample induces carriers to move from one defect state to another via hopping. The carriers hope for a longer range at lower frequencies due to the considerable time period available to them. In contrast, for higher frequencies, the small-time period provides the short-range motion of carriers. The temperature affects the hopping of carriers resulting in variation in conductivity \cite{Ortega_PRB_08, Kahouli_JPCA_12}. The calculated {\it a.c.} conductivity using the formula $\sigma$($\omega$)=$\omega$$\epsilon_0$$\epsilon_r$tan$\delta$ is shown in Fig.~\ref{ACS}(a) at different temperatures. It shows that the conductivity increases with frequency, which confirms the contribution from bound charge carriers. The {\it a.c.} conductivity variation with frequency shows two different regions: (i) a broad plateau-like region at lower frequencies, and (ii) a dispersed region at higher frequencies. The frequency dependence of {\it a.c.} conductivity is explained by the jump relaxation model given by Funke \cite{Funke_PSSC_93, Sumi_Jap_10}, which states that at lower frequencies the carriers undergo successful hopping for longer periods giving the long-range translational motion of carriers that provides the frequency-independent conductivity. For higher frequencies, two competing processes, i.e., unsuccessful and successful hopping occur due to carriers' shorter periods. The ratio of unsuccessful hopping to successful hopping creates the dispersed behavior in conductivity at higher frequencies. The frequency dependence of {\it a.c.} conductivity data are fitted using the Jonscher universal power-law given by \cite{Jonscher_Nature_77}
\begin{equation} 
\label{eqn-acsig}
	\sigma_{ac} (\omega) = \sigma_{dc} +A ~\omega^s
	\end{equation}
where $\sigma_{ac}$$(\omega)$ is total measured conductivity, $\sigma_{dc}$ is {\it d.c.} contribution (frequency-independent) and $A$$\omega$$^s$ describes the frequency-dependent contribution; here $A$ is the temperature-dependent constant that determines the amount or strength of polarization, and $s$ is an exponent representing the interaction between lattice and mobile ions depending on the temperature and frequency of the applied field. The changes from frequency-independent region to the frequency-dependent region show the transition from long-range hopping over barriers to the short-range motion of carriers \cite{Nasri_CerInt_16, Sharma_JMS_20}. The temperature dependence of the $s$ parameter provides the information about the type of conduction mechanism followed in {\it a.c.} conductivity \cite{Ortega_PRB_08}. For example, (a) if $s$ is nearly constant around 0.8 or increases slightly with temperature, the conduction mechanism followed is quantum mechanical tunnelling \cite{Ghosh_PRB_90}, (b) if $s$ decreases linearly with an increase in temperature, the conduction mechanism is governed by the correlated barrier hopping (CBH) \cite{Taher_MRB_16, Mollh_JAP_93}, (c) if $s$ increases linearly with an increase in temperature, the conduction mechanism is given by the non-overlapping small polaron hopping (NSPT) \cite{Ghosh1_PRB_90}, and (d) if $s$ depends on the frequency and temperature with decreases from unity at room temperature, reaches a minimum value and increases again with temperature, the conduction mechanism in this case is governed by overlapping large polaron tunnelling (OLPT) \cite{Kahouli_JPCA_12, Long_AIP_82}. 

Interestingly, the analysis of {\it a.c.} conductivity shows a similar type of conductivity behavior up to 360~K, and we observe an abrupt increase in conductivity at higher temperatures indicating the possibility of different conduction mechanisms as confirmed by the Jonscher power law. We have fitted the {\it a.c.} conductivity data using the Jonscher power law (equation $\ref{eqn-acsig}$) \cite{Jonscher_Nature_77} and the obtained parameters varies as $\sigma_{ac}$ between 2.4$\times$10$^{-6}$ and 1.4$\times$10$^{-4}$ S/m, $A$ between 1.9$\times$10$^{-11}$ and 6.4$\times$10$^{-7}$, and $s$ between 1.1 and 0.55 in the temperature range of 260--400~K. We found that the conductivity can not be fitted below 260~K due to strong frequency dependence. The temperature dependence of the $s$ parameter is plotted in Fig.~\ref{ACS}(b), which shows a decreasing trend at lower temperatures following the correlated barrier hopping (CBH) model within 260--360~K, and $s$ found to increase above 360~K following the non-overlapping small polaron tunneling (NSPT) model within the range of 370--400~K. In the CBH model, the conduction takes place between the localized defect states separated by Coulomb barrier of energy $W_M$ and the intersite separation $R_{\omega}$ is called as hopping distance. The decrease in hopping distance gives the potential overlapping of defects site leading to decrease in barrier height; hence, an increase in conductivity. The conduction in the CBH model is governed by either a single or bipolaron hopping \cite{Pike_PRB_72, Elliott_PM_77}, where the electron creates disorder in its surrounding medium by moving the atoms from their equilibrium positions gives the structural types of defects. In CBH model, $s$ parameter is given by \cite{Taher_MRB_16, Mollh_JAP_93}
\begin{equation} \label{s}
	s= 1-\frac{6k_{B}T}{W_{M} - k_{B}T (ln\frac {1}{\omega \tau_{0}})} 
	\end{equation}
where $W_M$ is the binding energy (Coloumb barrier height) of polaron to move from one atomic site to another. For the large values of $W_M$/$k_B$T,  the equation \ref{s} becomes 
\begin{equation} \label{eqn}
	s= 1-\frac{6k_{B}T}{W_{M}} 
	\end{equation}
From the above equation, the slope of (1$-s$) versus temperature (as shown in Fig.~\ref{ACS}(b)] gives the value of Coulomb barrier $W_M$ = 0.08~eV \cite{Brahim_MRB_19}. The hopping distance $R_{\omega}$ in the CBH model can be written as   
\begin{equation} \label{eqn}
	R_{\omega}= \frac {4n e^2}{\pi \epsilon_0 \epsilon_r  [W_{M} - k_{B}T (ln\frac {1}{\omega \tau_{0}})]}
	\end{equation}
where $n$ is the number of polarons involved in the hopping process ($n=$ 1 for single polaron hopping and $n=$ 2 for bi-polaron hopping). When $W_M\le$ $E_a$/4, the conduction mechanism is governed by single polaron hopping. In the present sample $W_M$ (=0.08~eV) found to be $\le$$E_a$/4 (0.17~eV), which suggests that the conduction is governed by single polaron hopping \cite{Brahim_MRB_19, Hajlaoui_PB_15}. 

\begin{figure} 
\includegraphics[width=3.4in]{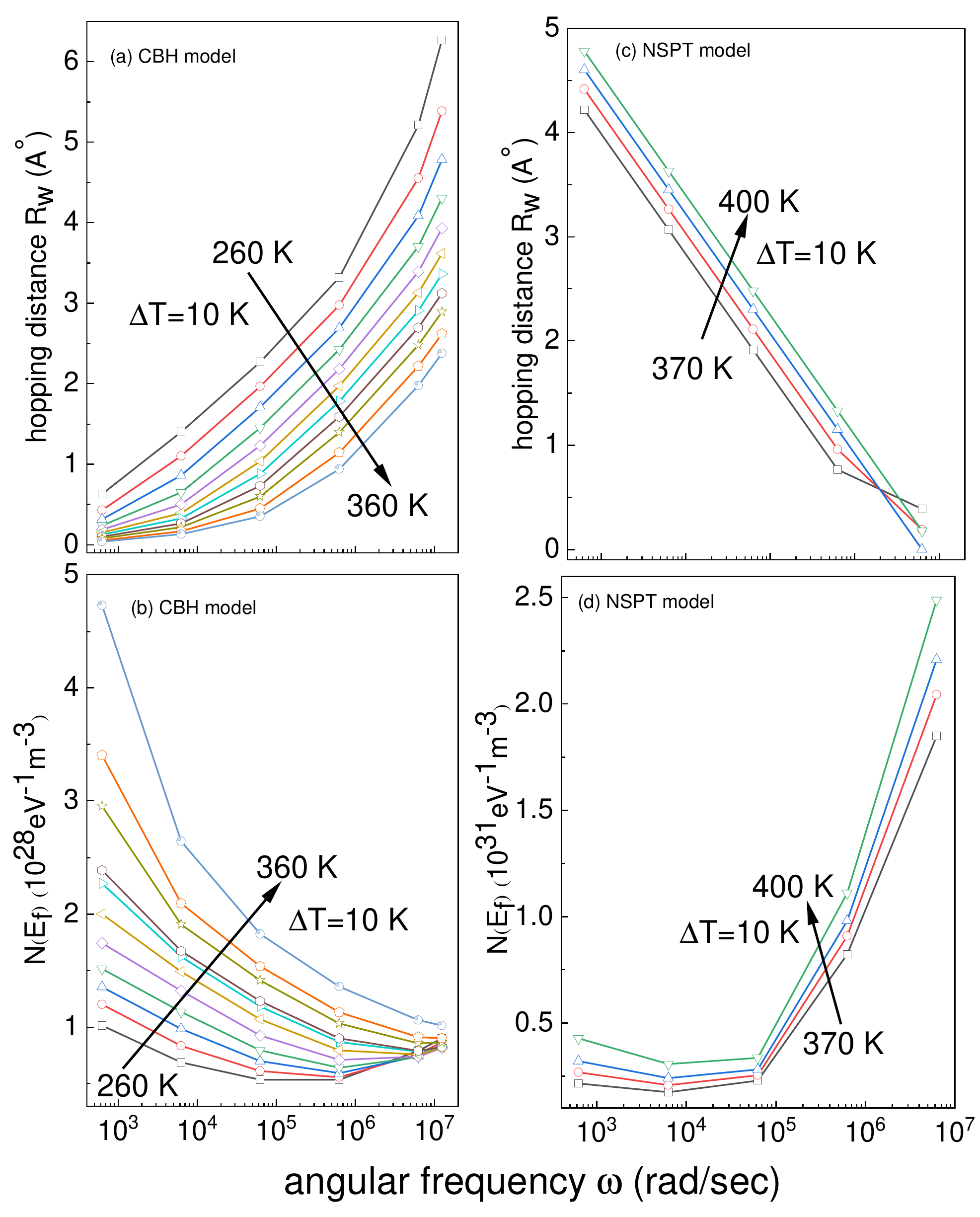}
\caption {The variation of hopping distances (a, c), and effective density of states (b, d) with frequency at different temperatures for the Na$_3$Zr$_2$Si$_2$PO$_{\rm 12}$ sample in CBH and NSPT regions, respectively.} 
\label{CNF}
\end{figure}

\begin{figure}[h] 
\includegraphics[width=3.45in]{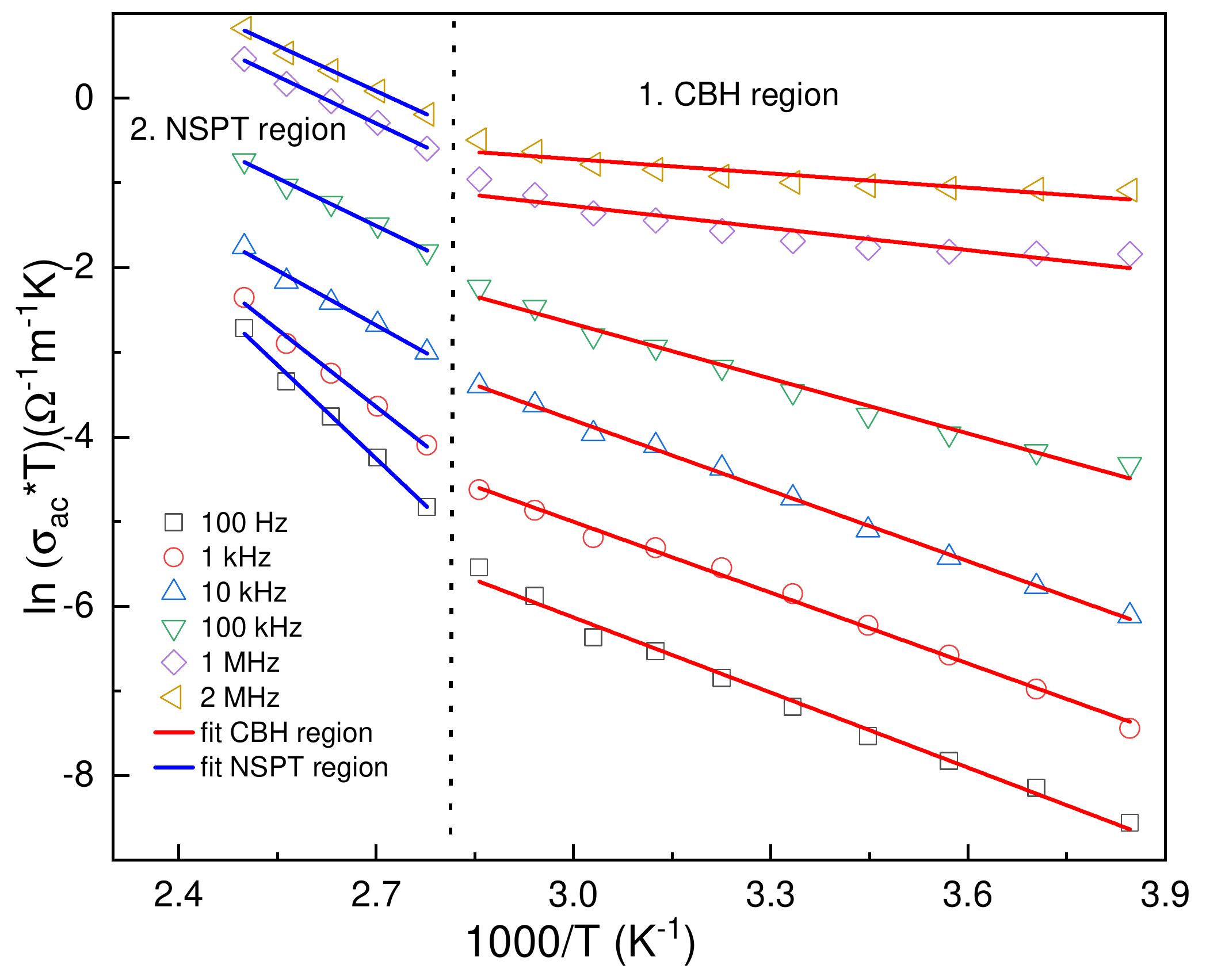}
\caption {The plot between ln($\sigma_{ac}$*T) vs. (1000/T) for the Na$_3$Zr$_2$Si$_2$PO$_{12}$ sample at selected frequencies, open symbols and solid lines show the experimental data and linear fit, respectively. The dashed line separates the regions of two different types of conduction mechanisms following NSPT and CBH models, respectively.} 
\label{sigma ACT}
\end{figure}

Moreover, the variation of hopping distance with frequency at different temperatures is shown in Fig.~\ref{CNF}(a). It is found that the hopping distance $R_{\omega}$ increases with frequency and decreases with temperature. Similarly, the density of states (DOS) near Fermi level in CBH region are calculated using Austin and Mott equation \cite{Mott_AIP_69} 
\begin{equation} 
\label{eqn-acCBH}
	\sigma_{ac} (\omega) =\frac {\pi}{3} e^2 k_B \alpha^ {-5} [N (E_F)]^2 \omega [ln \frac{1}{\omega\tau_{0}}]^4  T
	\end{equation}
where, $e$ is the electronic charge, and $\alpha$ is the spatial extension of the polaron ($\approx$$10^{10} m^{-1}$). The variation in DOS is shown in Fig.~\ref{CNF}(b), where the higher values (order of $10^{28}m^{-3}$) indicate that the charge transport is through the hopping over the localized defect states. The DOS increases with temperature, resulting in enhanced conductivity as more states are available for hopping conduction. The temperature dependence of the $s$ parameter in Fig.~\ref{ACS}(b) shows that the behavior/slope of $s$ (1-$s$) curves change abruptly above 360~K. The temperature dependence of the $s$ parameter suggests that the NSPT model is suitable to characterize the conduction mechanism in 370-400 K region. The polaron tunneling occurs in a covalent solid if the charge carrier causes a significant distortion in the lattice. The small polarons mean particles are so localized that the distorted lattice sites do not overlap and electrons can tunnel between the states near the Fermi level.  In the NSPT model, the $s$ parameter can be written as the follwoing \cite{Kahouli_JPCA_12, Mathlouthi_JALCOM_19}:
\begin{equation} \label{eqn}
	s= 1-\frac {4}{[ln\frac {1}{(\omega \tau_{0})}-(\frac{W_{H}}{k_{B}T})]}
	\end{equation}
where $W_H$ is the polaron binding energy. For the large values of $W_H$/$k_B$T, the above equation becomes
\begin{equation} \label{eqn}
	s= 1+\frac{4k_{B}T}{W_{H}} 
	\end{equation}
The slope of (1$-s$) versus temperature gives the binding energy of small polaron in the NSPT model, which found to be around 0.65~eV. The tunnelling distance (R$_{\omega}$) of polaron in the NSPT model is given by 
\begin{equation} \label{eqn}
R_{\omega}=\frac {1}{2\alpha}{[ln(\frac {1}{\omega \tau_{0}})-\frac{W_{H}}{k_{B}T}]}
\end{equation}
The variation in $R_{\omega}$ with temperature is shown in Fig.~\ref{CNF}(c), which found to decrease with frequency and increase with temperature. The obtained value of $R_{\omega}$ is found to be of the order of inter atomic spacing that confirms the hopping over the localized states. The density of states (DOS) near Fermi level in NSPT model are calculated using {\it a.c.} conductivity, as given by \cite{Kahouli_JPCA_12} 
\begin{equation}
\label{eqn}
	\sigma_{(ac)} \omega = \frac{\pi}{3} e^2{3} k_B T \omega \alpha^ {-1} [N (E_F)]^2 \frac {R_\omega^4} {12}
\end{equation}
where $N(E_F)$ is the DOS near the Fermi level. The variation in DOS with frequency at various temperatures is presented in Figs.~\ref{CNF}(d), which show that the DOS increases with temperature and having more sites available for the conduction, hence increase in conductivity. We note that the {\it a.c.} conductivity increases with temperature, as shown in Fig.~\ref{ACS}(a), which suggests for the thermally activated Arrhenius behavior, as given by $\sigma_{ac}$T = A $exp({-E_a}/{k_B T}$), where A is the pre-exponential factor and $E_a$ is activation energy. In order to validate this model, we plot ln($\sigma_{ac}$T) vs. (1000/T) at selected frequencies in Fig.~\ref{sigma ACT}. 
It offers two different conductivity regions (as separated by the dashed line) correspond to two different types of conduction mechanisms. The activation energy at various frequencies obtained by fitting the Arrhenius thermal conduction model varies between 0.25--0.5~eV (region 1) and 0.65--0.3~eV (region 2) in the frequency of 100~Hz--2~MHz. We find that the activation energy decreases with frequency because the enhanced hopping between localized states lowers the energy barrier and increase the conductivity.

\begin{figure} 
\includegraphics[width=3.4in]{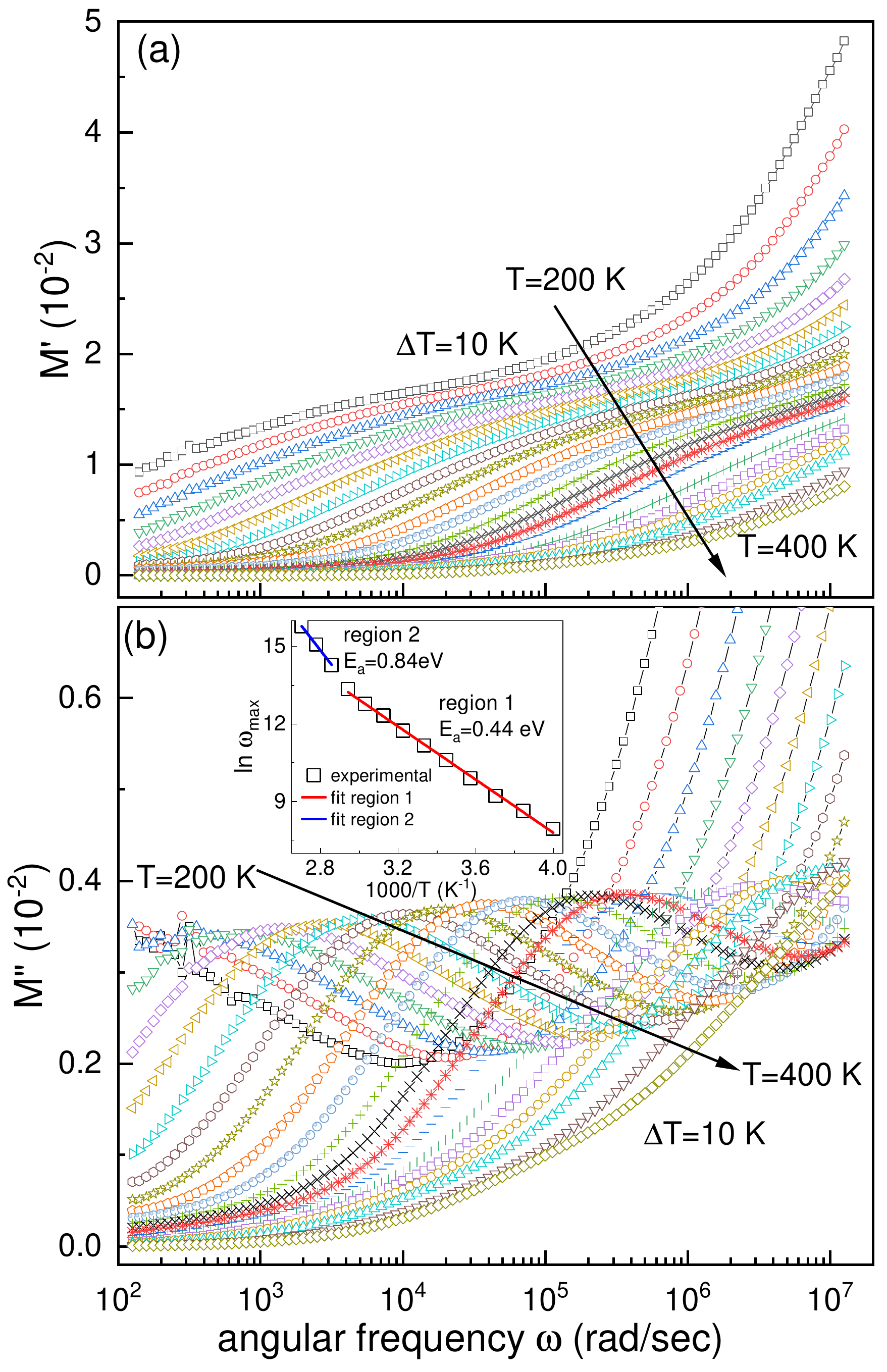}
\caption {The real (a) and imaginary (b) parts of the electric modulus as a function of angular frequency at various temperatures for the Na$_3$Zr$_2$Si$_2$PO$_{\rm 12}$ sample. The Arrhenius plot of ln($\omega_m$) versus 1000/T is shown in the inset of (b).}
\label{MMT}
\end{figure}

Finally we discuss the electric modulus to understand the relaxation phenomena in ionic conductors, whereas the electric displacement remains constant. The electrical modulus representation excludes the electrode polarization and separates the space-charge effect from bulk conductivity. The analysis of the real and imaginary components of the electric modulus provides information about the conduction mechanism, distribution of relaxation time, and charge carrier parameters of ionic conductors. \cite{Baskaran_JAP_02, Nasri_CerInt_16, Ksentini_APA_20}.  The electric modulus is defined as reciprocal of permittivity, represents an actual dielectric relaxation process, can be written as \cite{Moynihan_PCG_73, Hajlaoui_PB_15}: 
\begin{equation}
    \label{eqn}
    M^*(\omega)=M'(\omega)-jM''(\omega)=j \omega C_{0} Z^*
    \end{equation}
 where $Z^*= (Z'-jZ'')$, and the real $M'$ and imaginary $M''$ parts of complex electric modulus are given by
  \begin{equation}
    \label{eq-a}
      M'(\omega)=\omega C_0 Z'' \hskip0.5cm and \hskip0.5cm  M''(\omega)=\omega C_0 Z'
  \end{equation}
The real and imaginary components of electric modulus calculated using equation~\ref{eq-a} are shown in Figs.~\ref{MMT}(a, b), respectively. The smaller values of $M'$ at lower frequencies are due to the lack of restoring force that provides charge carriers mobility under the influence of an electric field. It is evident that the values of $M'$ approaching dispersed region towards $M_\infty$ [the asymptotic value of $M'(\omega)$] at higher frequency. The $M'$ decreases with temperature indicating the temperature-dependent relaxation process in the material \cite{Lakhdar_MSSP_15}. The $M''$ shown in Fig.~\ref{MMT}(b) is related to energy dissipation that occurs in irreversible conduction. It increases with frequency and attains a maxima at $\omega$$\tau$=1. The relaxation peak in modulus spectra shifts towards higher frequency with increasing the temperature, which indicates the thermally activated relaxation in the material. The electric modulus separates the motion of charge carriers from long range to short range. The frequency below $M''_{max}$ (peak maximum of the $M''$) shows the range where charge carriers are mobile over long distances (i.e., perform the long-range hopping of charge carriers from one site to another site), while the frequency range above the peak shows that the carriers are mobile over a short distance (confined within the potential walls) \cite{Nasri_CerInt_16}. 

The dielectric relaxation phenomenon is explained using the Laplace transformation of the Kohlraush-Williams-Watts (KWW) decay function given by $\phi$(t)=$\exp{(-\frac{t}{\tau})^\beta}$, where $\beta$ is stretched exponent varies between 0 and 1 decides the relaxation-time distribution behavior. The $\beta=1$ corresponds to the ideal Debye type behavior where dipole-dipole interactions are negligible, whereas the $\beta=0$ suggests for the maximum interaction among dipoles. The KWW function relates the decay function with the relaxation time $\tau$ and provide the information about time dependence of the electric field within the dielectric material \cite{Moynihan_PCG_73}. The KWW function can be modified to have the analysis in the frequency domain \cite{Bergman_JAP_00, Baskaran_JAP_02, Ponpandian_JAPD_02}, where the imaginary component of electric modulus can be written as
\begin{equation}
\label{eq-M}
    M''=\frac{M''_{max}}{[(1-\beta)+(\frac{\beta}{1+\beta})[\beta(\frac{\omega_m}{\omega})+(\frac{\omega}{\omega_m})^\beta]}
\end{equation}
 The value of $\beta$ calculated using the equation \ref{eq-M} shows that the values vary from 0.33 to 0.17, which found to be less than unity and that confirms the non-Debye type relaxation behavior in the materials over the measured temperature range. The activation energy of the most probable relaxation frequency can be found using the Arrhenius law, which follows the temperature dependence as given by the following equation:
\begin{equation}
    \label{eq-Arr}
    \omega_{m} = \omega_0 ~ exp [\frac{-E_a}{k_BT}]
\end{equation}
where $\omega_0$ is the pre-exponential factor and $E_a$ is the activation energy of the relaxation. The activation energy is calculated from the linear fit of ln$(\omega_{m})$ versus 1000/T plot, which found to be 0.44~eV in region 1 and 0.84~eV in region 2, as shown in the inset of Fig.~\ref{MMT}(b). These two activation energies are correspond to two different type of relaxation mechanisms followed in the material.

\begin{figure} [h]
\includegraphics[width=3.4in]{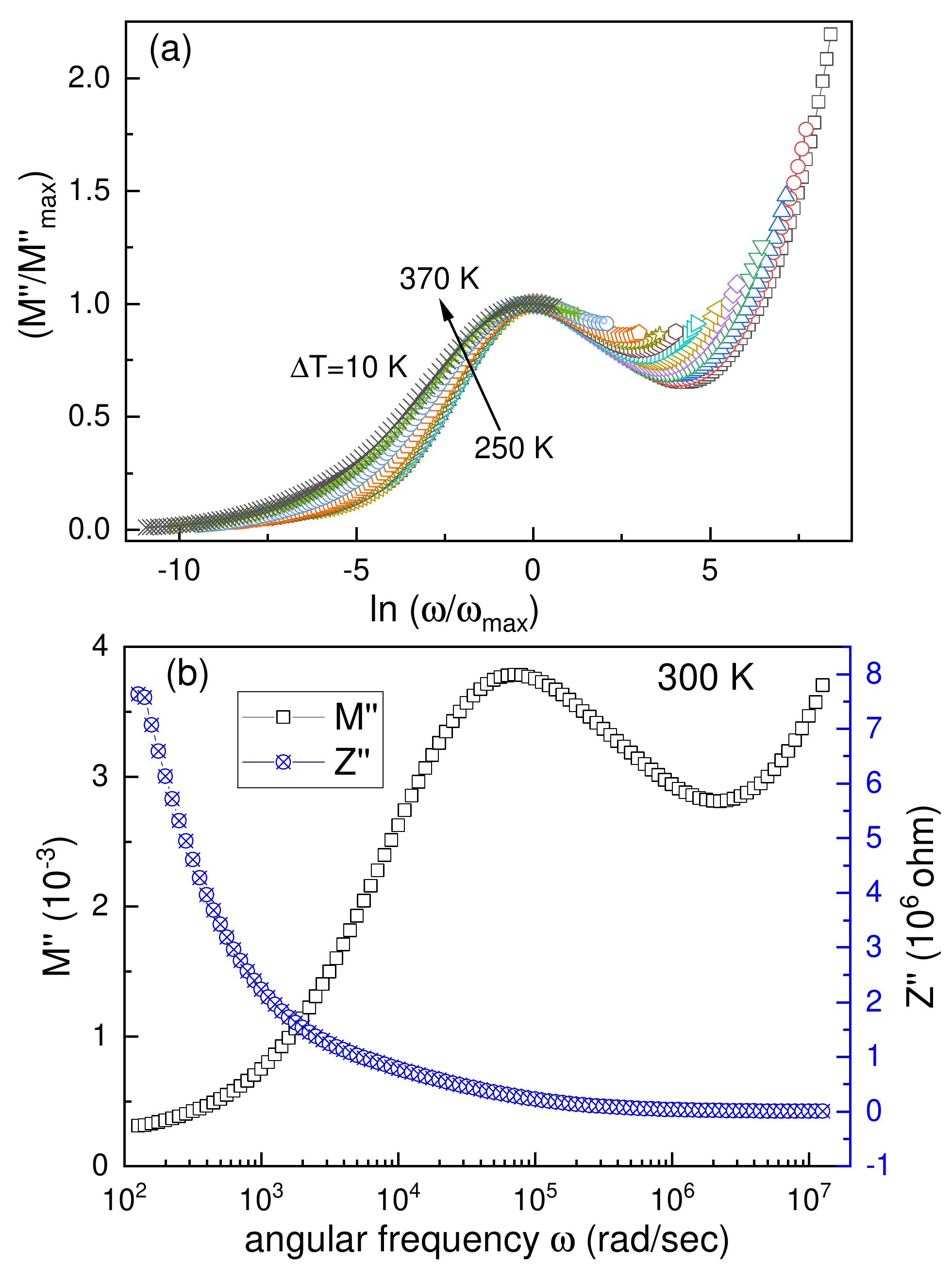}
\caption {(a) Scaling behavior of imaginary part of electric modulus $M''$ versus angular frequency $\omega$ at various temperature. (b) The combined impedance and electric modulus spectrum of Na$3$Zr$_2$Si$_2$PO$_{\rm 12}$ sample at room temperature.}
\label{SMZ}
\end{figure}

The scaling behavior of the electric modulus ($M''$/$M''_{max}$) vs. ln($\omega$/$\omega_{max}$) is shown in Fig.~\ref{SMZ}(a) where $M''_{max}$ is peak value of $M''$ and $\omega_{max}$ is the frequency corresponding to the peak in electric modulus. The analysis shows that peaks between 250--370~K tending to merge into a single curve, which confirms the same type of relaxation mechanism in this temperature range. Also, the tendency of merging all the curves in a single peak shows the dynamic carrier process that occurs at different time scales and gives the similar type thermal activation energy. The combined analysis of $M''$ and $Z''$ with frequency are useful to understand the type of relaxation phenomena in the material, and the kind of mobility (long range or short range) of charge carriers, as shown in Fig.~\ref{SMZ}(b). The amplitude of $Z''$ is proportional to the most resistive element, while the amplitude of $M''$ depends on the lowest capacitive element. If the peaks in $Z''$ and $M''$ coincide at the same angular frequency, the mobility of the ions is due to the long-range mobility of charge carriers, and relaxation is of the Debye type. If the peak of both the curves doesn't occur at the same angular frequency, the mobility is due to the short-range mobility of charge carriers.  As seen in Fig.~\ref{SMZ}(b) that both the graphs of $Z''$ and $M''$ do not coincide with each other, confirming the short-range mobility of charge carriers with non Debye type wide distribution of relaxation times. Similar type behavior is also observed in several other materials at different measured temperatures \cite{Hajlaoui_PB_15, Ksentini_APA_20, R.Gerhardt_JPCS_94}.

\section{\noindent ~Conclusions}

In summary, we have investigated the physical properties of Na$_3$Zr$_2$Si$_2$PO$_{\rm 12}$ bulk sample prepared by solid-state reaction method. The Rietveld refinement show the monoclinic phase formation with space group (C 2/c). The {\it d.c.} resistivity measurements following the Arrhenius thermal conduction model with activation energy of 0.68~eV. The temperature dependence of dielectric constant and dielectric loss are explained using the space charge polarization mechanism. The dielectric loss peak follows the thermally activated relaxation process with an activation energy of 0.25~eV. The decrease in impedance with frequency and temperature is explained by the reduction in barrier height and space charge polarization. The {\it a.c.} conductivity data fitted using the Jonschers power law shows two different conduction mechanisms, (a) correlated barrier hopping and (b) non-overlapping small polaron tunneling model in the measured temperature range. The conduction  due to the single polaron is confirmed by the CBH model. The Arrhenius fits of {\it a.c.} conductivity at various frequencies show the increase due to the reduction in activation energy. The electric modulus study shows the thermally activated non-Debye type relaxation behavior. The combined analysis of the imaginary part of electric modulus and impedance shows the short-range mobility of charge carriers. 

\section{\noindent ~Acknowledgments}

RM  thanks IUAC for providing the experimental facilities. We thank Ajay Kumar and Vikas Thakur for help and useful comments/suggestions. RSD acknowledges the financial support from SERB-DST through a core research grant (project reference no. CRG/2020/003436). 






\end{document}